\documentclass[prc,showpacs,nofootinbib,preprint]{revtex4-1}
\usepackage{epsfig,graphics,color}
\usepackage{subfigure}
\usepackage{bbm}
\usepackage{amssymb}
\usepackage{amsmath}

\newcommand{\bra}[1]{\langle #1|}
\newcommand{\ket}[1]{|#1\rangle}

\newcommand{\roundbra}[1]{(#1|}
\newcommand{\roundket}[1]{|#1)}

\def\one{{\bf 1}\,}
\def\quarknumberoperator{{\mathbbm 1}\,}

\def\tr{{\rm tr} \,}

\def\vslash{v \hspace{-1.7mm}/}

\def\w2{\tilde w^2}
\def\ws2{1}

\newcommand{\smallfrac}[2]{\textstyle{\frac #1 #2}}

\begin{document}
\title{ Combined large-$N_c$ and heavy-quark operator analysis \\ of 2-body meson-baryon
counterterms \\ in the chiral Lagrangian with charmed mesons}
\author{M.F.M. Lutz$^1$, D. Samart$^{1,2}$ and A. Semke$^1$}
\affiliation{$^1$ GSI Helmholtzzentrum f\"ur Schwerionenforschung GmbH,\\
Planck Str. 1, 64291 Darmstadt, Germany}
\affiliation{$^2$ Suranaree University of Technology, Nakhon Ratchasima, 30000, Thailand}
\date{\today}
\begin{abstract}
The chiral $SU(3)$ Lagrangian with pseudoscalar and vector $D$ mesons and with the octet and decuplet baryons is
considered. The leading two-body counter terms involving two baryon fields and two $D$ meson fields are constructed
in the open-charm sector. There are 26 terms. A combined expansion in the inverse of the charm quark mass and in the
inverse of the number of colors provides sum rules that reduce the number of free parameter down to
5 only. Our result shows the feasibility of a systematic computation of the open-charm baryon spectrum based on
coupled-channel dynamics.
\end{abstract}

\maketitle

\section{Introduction}

The chiral $SU(3)$ Lagrangian with the pseudoscalar and vector $D$ mesons has been applied extensively in the
literature (see e.g. \cite{Georgi:1990,Wise92,Casalbuoni,Mehen-Springer-2004}).
Most exciting are recent studies on s-wave  scattering of Goldstone bosons off
$D$ mesons \cite{Kolomeitsev-Lutz-2004,Hofmann-Lutz-2004,Lutz-Soyeur-2006,Guo-Hanhart-Krewald-Meissner-2008}
The leading order coupled-channel interaction leads to the formation of scalar and axial-vector $D$ mesons with
properties compatible with the known empirical constraints. Such results are similar to findings on the s-wave scattering
of the Goldstone bosons off the baryon octet and decuplet states
(see e.g. \cite{Kaiser-Siegel-Weise-1995,Oller-Meissner-2001,Kolomeitsev-Lutz-2004-Baryon}), where the
coupled-channel dynamics based on the
leading order chiral Lagrangian generates s- and d-wave baryon resonances.

The main objective of the present study is to pave the way for systematic coupled-channel computations on open-charm
baryon resonances. A first application of the chiral Lagrangian to s-wave baryon resonances considered the
coupled-channel interaction of the Goldstone bosons with the ground-state baryons with open
charm \cite{Lutz-Kolomeitsev-2004-charm}. A rich spectrum of
chiral excitations was obtained. The challenge of the open-charm sector is the possibility of charm-exchange reactions,
where the charm of the baryon is transferred to the meson.  A first phenomenological case study modeled the
coupled-channel force by a t-channel exchange of vector mesons in the
static limit \cite{Hofmann-Lutz-2005,Hofmann-Lutz-2006,Tolos-2004}. Such a t-channel
force recovers the leading order predictions of the chiral symmetry whenever Goldstone bosons are involved. It was shown
in \cite{Hofmann-Lutz-2005,Hofmann-Lutz-2006,Ramos-Vidana-2009} that a rich spectrum of s- and d-wave baryon resonances is generated dynamically based on such a schematic ansatz. Two types of resonances are generated. The resonances of the first type are formed predominantly by the interaction of the
Goldstone bosons with the open-charm baryon ground states, and the second type is a consequence of the coupled-channel
interaction of the $D$ mesons with the octet and decuplet baryons. For both types of resonances the exchange of the
light vector mesons constitute the driving forces that generate the hadronic molecules. The exchange
of charmed vector mesons leads to the  typically small widths of the second type resonances.

While the interaction of the Goldstone bosons with any hadron is dictated by 
chiral symmetry, this is not true for the interaction of the $D$ mesons with e.g. baryons. At leading order the $D$ mesons
interact with baryons via local counter terms that are undetermined by chiral symmetry. This resembles the situation
encountered  in chiral studies of the nucleon-nucleon force (for a review see e.g. \cite{Epelbaum-Hammer-Meissner-2009}).
Only the long-range part of the interaction is controlled
by chiral interactions, the short range part needs to be parameterized in terms of a priori unknown contact terms.
Needless to state that a reliable coupled-channel computation requires the consideration of both, the short-range and
long-range forces.

The purpose of the present work is a systematic construction of the leading order contact terms
for the interaction of the $D$ mesons with the baryon octet and decuplet states. Though the chiral symmetry does not constraint such contact terms,
 there are significant correlations amongst the counter terms implied by the
heavy-quark symmetry and the large-$N_c$ limit of QCD. In the limit of a large charm quark mass the pseudoscalar and vector
$D$ mesons form a spin multiplet with degenerate properties \cite{Georgi:1990}. Thus, a systematic coupled-channel
computation in this limit requires the simultaneous consideration of pseudoscalar and vector $D$ mesons. This leads necessarily
to the presence of long-range t-channel forces. The transition of a vector $D$ meson to a pseudoscalar $D$ meson involves
a Goldstone boson.  A first attempt to consider pseudoscalar and vector $D$ mesons on an equal footing, however,
assuming zero range forces only, can be found in \cite{Gammermann-2010}.
Similarly, in the limit of a large number of colors in QCD the baryon octet and
decuplet states form a super multiplet \cite{Gervais1983, Dashen:Jenkins:Manohar:1994:1}. This asks for the simultaneous consideration of
the octet and decuplet baryons.

In a recent study two of the authors worked out the implications of large-$N_c$ QCD on the local counter terms of the
Goldstone boson interaction with the baryon octet and decuplet states \cite{Lutz-Semke-2011}.  The technology developed in
\cite{Luty:Russel:1994,Dashen:Jenkins:Manohar:1994} was applied. Matrix elements
of current-current correlation functions were evaluated in the baryon octet and decuplet states. The matrix elements
were expanded in powers of $1/N_c$ applying the operator reduction rules of \cite{Dashen:Jenkins:Manohar:1994}.
This technology is well suited for an application to the open-charm sector.
The implications of heavy-quark symmetry on the counter terms can be worked out using a suitable multiplet
representation of the pseudoscalar and vector $D$ mesons \cite{Wise92,Casalbuoni,Mehen-Springer-2004}.

\section{Chiral Lagrangian with meson and baryon fields} \label{section:chiral-lagrangian}

The construction rules for the chiral $SU(3)$ Lagrangian density are recalled.
For more technical details see for example \cite{Weinberg68,GL84,Krause:1990,Kaym84,Ecker89,Borasoy,Birse,Becher,Fuchs}.
The basic building blocks of the chiral Lagrangian  are
\begin{eqnarray}
&& U_\mu = {\textstyle{1\over 2}}\,e^{-i\,\frac{\Phi}{2\,f}} \left(
\partial_\mu \,e^{i\,\frac{\Phi}{f}} \right) e^{-i\,\frac{\Phi}{2\,f}} \;, \qquad  B\,, \qquad B_\mu\,, \qquad
D\,, \qquad D_{\mu \nu} \,,
\nonumber\\
&& \chi_\pm = \frac{1}{2} \left(
e^{+i\,\frac{\Phi}{2\,f}} \,\chi_0 \,e^{+i\,\frac{\Phi}{2\,f}}
\pm e^{-i\,\frac{\Phi}{2\,f}} \,\chi_0 \,e^{-i\,\frac{\Phi}{2\,f}}
\right) \,,
\label{def-fields}
\end{eqnarray}
where we include the pseudoscalar meson octet fields
$\Phi(J^P\!\!=\!0^-)$,
the baryon octet fields $B(J^P\!\!=\!{\textstyle{1\over2}}^+)$ and the baryon
decuplet fields $B_\mu(J^P\!\!=\!{\textstyle{3\over2}}^+)$.  Furthermore we introduce the $SU(3)$ flavor antitriplet
fields $D(J^P\!\!=\!0^-)$ and $D_{\mu \nu}(J^P\!\!=\!1^-)$
to describe the pseudoscalar and vector
$D$ mesons \cite{Lutz-Soyeur-2006}. Explicit chiral symmetry-breaking
is included in terms of scalar and pseudoscalar source fields $\chi_\pm $ proportional to the quark mass
matrix of QCD with $\chi_0 =2\,B_0\, {\rm diag} (m_u,m_d,m_s)$.
The merit of the particular field combinations in Eq.~(\ref{def-fields})  is their
identical transformation property under chiral $SU_L(3)\otimes SU_R(3)$ rotations.

The octet fields may be decomposed into their isospin multiplets,
\begin{eqnarray}
&& \Phi = \tau \cdot \pi
+ \alpha^\dagger \cdot  K +  K^\dagger \cdot \alpha
+ \eta\,\lambda_8\,,
\nonumber\\
&& \sqrt{2}\,B =  \alpha^\dagger \!\cdot \! N + \lambda_8 \,\Lambda + \tau \cdot \Sigma
 +\Xi^t\,i\,\sigma_2 \!\cdot \!\alpha   \, ,
\nonumber\\
&&\; \alpha^\dagger = {\textstyle{1\over \sqrt{2}}}\,(\lambda_4+i\,\lambda_5 ,\lambda_6+i\,\lambda_7 )\,,\qquad
\tau = (\lambda_1,\lambda_2, \lambda_3)\,,
\label{def-octet-fields}
\end{eqnarray}
with $K=(K^{(+)},K^{(0)})^t$,  for instance. The matrices $\lambda_i$ are the Gell-Mann matrices and $\sigma_2$ is
the second Pauli matrix. The open-charm fields $D$ and $D_{\mu\nu}$ form flavor antitriplets with e.g.
$D=(D^0,-D^+,D^+_s) $. The baryon decuplet fields $B^{abc}_\mu $ are completely symmetric in $a,b,c=1,2,3$ and
are related to the physical states by
\begin{eqnarray}
\begin{array}{llll}
B^{111} = \Delta^{++}\,, & B^{113} =\Sigma^{+}/\sqrt{3}\,, &
B^{133}=\Xi^0/\sqrt{3}\,,  &B^{333}= \Omega^-\,, \\
B^{112} =\Delta^{+}/\sqrt{3}\,, & B^{123} =\Sigma^{0}/\sqrt{6}\,, &
B^{233}=\Xi^-/\sqrt{3}\,, & \\
B^{122} =\Delta^{0}/\sqrt{3}\,, & B^{223} =\Sigma^{-}/\sqrt{3}\,, &
& \\
B^{222} =\Delta^{-}\,. & & &
\end{array}
\label{def-decuplet-fields}
\end{eqnarray}
To cope with flavor indices in the products of SU(3) tensors containing decuplet fields, we shall apply a
compact \textit{dot-notation}
suggested in \cite{Lutz:Kolomeitsev:2002}:
\begin{eqnarray}\label{def:dot-notation}
(\bar B^\mu \cdot B_\nu)^m_k \equiv \bar B^\mu_{ijk} \,B_\nu^{ijm}, \quad (\bar B^\mu \cdot \Phi)^m_k \equiv \bar B^\mu_{ijk}\, \Phi^i_l\, \epsilon^{jlm}, \quad (\Phi \cdot B_\mu)^m_k \equiv B_\mu^{ijm}\, \Phi_i^l\, \epsilon_{jlk}\,.
\end{eqnarray}

The chiral Lagrangian consists of all possible interaction
terms, formed with the fields $U_\mu,\, B,\, B_\mu, \, D\,, D_{\mu \nu} $ and $\chi_\pm$.
Derivatives of the fields must be included in compliance with the local chiral $SU(3)$ symmetry. This leads to
the notion of a covariant derivative ${\mathcal D}_\mu$ which is identical for all fields in Eq.~(\ref{def-fields}).
For baryons, the covariant derivative ${\mathcal D}_\mu$  acts on the octet and decuplet fields
as follows:
\begin{eqnarray}
&&({\mathcal D}_\mu  B)^{a}_{b} = \partial_\mu B^{a}_{b} +  \Gamma^{a}_{\mu,l}\, B^{l}_{b} -
\Gamma^{l}_{\mu,b}\, B^{a}_{l} \,, \quad
\nonumber\\
&& ({\mathcal D}_\mu B_\nu)^{abc} = \partial_\mu B^{abc}_\nu + \Gamma^{a}_{\mu,l } B^{lbc}_{\nu}
+ \Gamma^{b}_{\mu,l } B^{alc}_{\nu}+\Gamma^{c}_{\mu,l } B^{abl}_{\nu} \,,
\label{def-covariant-derivative}
\end{eqnarray}
with $\Gamma_\mu=-\Gamma_\mu^\dagger$ given by
\begin{eqnarray*}
 &&\Gamma_\mu ={\textstyle{1\over 2}}\,e^{-i\,\frac{\Phi}{2\,f}} \,
\Big[\partial_\mu -i\,(v_\mu + a_\mu) \Big] \,e^{+i\,\frac{\Phi}{2\,f}}
+{\textstyle{1\over 2}}\, e^{+i\,\frac{\Phi}{2\,f}} \,
\Big[\partial_\mu -i\,(v_\mu - a_\mu)\Big] \,e^{-i\,\frac{\Phi}{2\,f}}\,.
\end{eqnarray*}
Analogous expressions hold for the covariant derivatives for the open-charm fields.

The chiral Lagrangian is a powerful tool, once it is combined with appropriate
counting rules leading to a systematic approximation strategy.
We aim at describing hadronic interactions at low-energy by constructing an expansion
in small momenta and small quark masses. In the following we construct the leading order
two-body counter terms involving two $D$ meson and two baryon fields.

We begin with the terms with two baryon octet fields. There are 12 leading order terms
\begin{eqnarray}
&& {\mathcal L} = D\,\Big\{c_1^{(S)}  (\bar B\,B)_+ + c_2^{(S)}  (\bar B\,B)_- + \frac 12\, c_3^{(S)} {\rm tr}\,(\bar B\,B) \Big\}\,\bar D \nonumber \\
&& \qquad - \frac{1}{2}\,D_{\mu \nu}\,\Big\{\tilde c_1^{(S)}  (\bar B\,B)_+ + \tilde c_2^{(S)} (\bar B\,B)_- + \frac 12 \,\tilde c_3^{(S)} {\rm tr}\,(\bar B\,B) \Big\}
\,\bar D^{\mu\nu}
\nonumber\\
&& \qquad + \, \frac{i}{M_c}\,D_{\mu \nu}\,\Big\{
c_1^{(A)} (\bar B\,\gamma_5\, \gamma^\mu\,B)_+  +  c_2^{(A)} (\bar B\,\gamma_5\, \gamma^\mu\,B)_- + \frac 12 \,c_3^{(A)}\,{\rm tr} \,(\bar B\,\gamma_5\, \gamma^\mu\,B)\Big\} \,(\partial^\nu \bar D)
 + {\rm h.c.}
\nonumber\\
&& \qquad + \,\frac{1}{4 \,M_c} \epsilon^{\mu \nu \alpha \beta}
D_{\mu \nu} \Big\{
\tilde c_1^{(A)} (\bar B \,\gamma_5\, \gamma_\alpha \,B)_+ + \tilde c_2^{(A)} (\bar B \,\gamma_5 \,\gamma_\alpha \,B)_- 
\nonumber\\
&& \qquad \qquad \qquad \qquad \quad +\, \frac 12 \,\tilde c_3^{(A)} {\rm tr} \,(\bar B \,\gamma_5\, \gamma_\alpha \,B)\Big\}
(\partial^\tau \bar D_{\tau \beta} ) + {\rm h.c.}\,, 
\label{def-La}
\end{eqnarray}
where $\bar D = D^\dagger$, $M_c$ is the mass of the charm quark and $(\bar B \,\Gamma\, B)_\pm$ stands for (anti-)\-commutator in flavor space. All structures describe s-wave scattering and
therefore we assign the chiral power $Q^0$ to them. The total number of terms
in Eq.~(\ref{def-La}) is readily understood. There are three different flavor $SU(3)$ invariants only implying three types of coupling constants
$c_1^{(\cdots)}, c_2^{(\cdots)} $ and $c_3^{(\cdots)}$. It remains to understand the four spin structures.
There is one spin structure involving the pseudoscalar $D$ mesons with the coupling constants $c_{\ldots}^{(S)}$.
This follows since a two-body system of a spin zero and a spin one-half particle allows to form one s-wave state only.
In contrast, two s-wave states  involving the vector $D$ mesons are possible. The relevant coupling constants are
introduced with $\tilde c_{\ldots}^{(S)}$ and $\tilde c_{\ldots}^{(A)}$.  The terms parameterized with $c_{\ldots}^{(A)}$
describe the s-wave transitions from a pseudoscalar to a vector $D$ meson.

We continue with the leading order terms involving two baryon decuplet fields.
At leading order we find the relevance of the following 10 terms
\begin{eqnarray}
&& {\mathcal L} = -D\,\Big\{d_1^{(S)} \,\bar B^\sigma \cdot B_\sigma
+ \frac{1}{2}\,d_2^{(S)} {\rm tr}\,(\bar B^\sigma \cdot B_\sigma) \Big\}\,\bar D
\nonumber\\
&& \qquad + \frac{1}{2}\,D_{\mu \nu}\,\Big\{\tilde d_1^{(S)} \,\bar B^\sigma \cdot B_\sigma
+ \frac{1}{2}\,\tilde d_2^{(S)} {\rm tr}\,(\bar B^\sigma \cdot B_\sigma) \Big\}
\,\bar D^{\mu\nu}
\nonumber\\
&& \qquad + \, \frac{i}{4\, }\,\epsilon^{\mu \nu \alpha \beta }\,D_{\mu \nu}\,\Big\{
d_1^{(E)}\,\bar B_\alpha\,\cdot B_\beta
+\frac{1}{2}\,d_2^{(E)}\,{\rm tr} \,(\bar B_\alpha  \cdot B_\beta)\Big\} \,\bar D
 + {\rm h.c.}
\nonumber\\
&& \qquad + \frac{1}{2}\,D_{\beta \mu}\,\Big\{\tilde d_1^{(E)} \,\bar B_\alpha\, \cdot B^\beta
+ \frac{1}{2}\,\tilde d_2^{(E)} {\rm tr}\,(\bar B_\alpha \cdot B^\beta) \Big\}
\,\bar D^{\alpha \mu}\nonumber\\
&& \qquad - \frac{1}{2}\,D_{\alpha \mu}\,\Big\{\tilde d_3^{(E)} \,\bar B_\alpha\, \cdot B^\beta
+ \frac{1}{2}\,\tilde d_4^{(E)} {\rm tr}\,(\bar B_\alpha \cdot B^\beta) \Big\}
\,\bar D^{\beta \mu}\,.
\label{def-Lb}
\end{eqnarray}
In this case there are two independent flavor structures. Again, there is one spin structure involving two pseudoscalar $D$ meson fields with coupling constants $d_{\ldots}^{(S)}$.
Since a two-body system of a spin three-half and a spin one  particle allows to form three s-wave states only,
there are three different contractions of the Lorentz indices in the terms containing two vector meson fields.
There remains the last spin structure parameterized by $d^{(E)}_{\ldots}$
describing the s-wave transitions from a pseudoscalar to a vector $D$ meson.

We close the collection of our counter terms with structures involving a baryon octet and a decuplet field.
At leading order we find the following 4 terms
\begin{eqnarray}
&&{\mathcal L}= \frac{i}{4 }\,\epsilon^{\mu \nu \alpha \beta}\Big\{
e^{(A)}_1\, D\,(\bar B_\mu \cdot\gamma_5\,\gamma_{\nu}\,B) \, \bar D_{\alpha \beta}) +   e^{(A)}_2\,D_{\alpha \beta} \,
(\bar B_\mu \cdot \gamma_5\,\gamma_{\nu}\,B) \, \bar D + {\rm h.c.}\Big\}
\nonumber\\
&& \qquad + \, \frac{\tilde  e^{(A)}_1}{2 }\,  D^{\tau \nu} \,
(\bar B_\mu \cdot \gamma_5\,\gamma_{\nu}\,B) \,\bar D_{\tau}^{\;\, \mu}  + {\rm h.c.}
\nonumber\\
&& \qquad - \, \frac{\tilde  e^{(A)}_2}{2 }\,  D^{\tau \mu} \,
(\bar B_\mu \cdot \gamma_5\,\gamma_{\nu}\,B) \,\bar D_{\tau}^{\;\, \nu}  + {\rm h.c.}  \,.
\label{def-Lc}
\end{eqnarray}

Altogether we constructed 26 leading order two-body counter terms. In the following we shall use the
heavy-quark symmetry in order to correlate the coupling constants introduced above.

\section{Heavy quark mass expansion}\label{section:HQS}

In order to work out the implications of the heavy-quark symmetry of QCD it is useful to introduce auxiliary and
slowly varying fields, $P_\pm(x)$ and $P^\mu_\pm(x)$. We decompose the $D$ meson fields into such fields
\begin{eqnarray}
&& D(x) \;\;\,\,\!= e^{-i\,(v\cdot x) \,M_c}\,P_{+}(x) +e^{+i\,(v\cdot x) \,M_c}\,P_{-}(x)\,,
\nonumber\\
&& D^{\mu \nu}(x)=
i\,e^{-i\,(v\cdot x) \,M_c}\,\Big\{v^\mu\,P^\nu_{+}(x)-v^\nu \,P^\mu_{+}(x)
+ \frac{i}{M_c}\,\Big( \partial^\mu P^\nu_+-\partial^\nu P^\mu_+\Big)\Big\}
\nonumber\\
&& \qquad \quad \;\,+\,i\,e^{+i\,(v\cdot x) \,M_c}\,\Big\{v^\mu\,P^\nu_{-}(x)-v^\nu \,P^\mu_{-}(x)
- \frac{i}{M_c}\,\Big( \partial^\mu P^\nu_--\partial^\nu P^\mu_-\Big)\Big\}\,,
\label{non-relativistic-expansion}
\end{eqnarray}
with a 4-velocity $v$ normalized by $v^2=1$. The mass parameter $M_c$ was already introduced in Eq.~(\ref{def-La}).
As a consequence of Eq.~(\ref{non-relativistic-expansion}), time and spatial derivatives of the fields $\partial_\alpha P_\pm$
are small compared to $M_c\,v_\alpha\,P$. In the limit $M_c \to \infty$ the former
terms can be neglected. Note that the fields $P_\pm$ and $P_\pm^\mu$
annihilate quanta with charm content $\pm 1$.
>From the equation of motion for the vector $D$ mesons it follows
\begin{eqnarray}
&& \partial^\mu \partial_\alpha D^{\alpha \nu}-\partial^\nu \partial_\alpha D^{\alpha \mu} +M_c^2\,D^{\mu \nu} =0
\qquad \leftrightarrow \qquad
\nonumber\\ \nonumber\\
&&\big\{(\partial \cdot \partial) \mp 2\,M_c\,i\,( v \cdot \partial )\big\}\,P^\mu_\pm = 0  \quad \& \quad
\mp i\,M_c\,v_\mu \, P^\mu_\pm + \partial_\mu \, P^\mu_\pm =0\,,
\label{rewrite-eom-tensor}
\end{eqnarray}
which teaches us that any term $v_\mu P^\mu $ vanishes in the heavy-quark mass limit.

In the limit of infinite quark mass, the fields $P_\pm$ and $P_\pm^\mu$ may be combined into an appropriate
multiplet field. This reflects the fact that in this limit the $1^-$ and $0^-$ mesons are related by a spin
flip of the charm quark, which does not cost any energy. Therefore, the properties of pseudoscalar and
vector states should be closely related. We follow here the formalism developed in
\cite{Wise92,YCCLLY92,BD92,Jenkins94,Casalbuoni} and introduce the multiplet field $H$, connected to the fields $P_+$ and $P_+^\mu$ as follows\footnote{Note that
${\tr } \gamma_5\,\gamma_\mu \,\gamma_\nu \,\gamma_\alpha \,\gamma_\beta =
-4\,i\,\epsilon_{\mu \nu \alpha \beta }
$
in the convention used in this work.}
\begin{eqnarray}
&& H = \frac{1}{2}\,\Big( 1 + \vslash \Big)\,\Big(\gamma_\mu\,P^\mu_+  +i\, \gamma_5\,P_+ \Big)
\nonumber\\
&& \bar H = \gamma_0\,H^\dagger\,\gamma_0 =
\Big(P^\dagger_{+,\mu} \,\gamma^\mu +P^\dagger_+ \,i\, \gamma_5 \Big)\,\frac{1}{2}\,\Big( 1 + \vslash \Big) \,,
\nonumber\\
&& P^\mu_+ \,v_\mu =0 \, , \qquad  v^2=1\,,
\label{phase-convention}
\end{eqnarray}
in terms of which the interaction should be composed. According to Ref.~\cite{Georgi:1990}, the field $H$ transforms
under the heavy-quark spin symmetry group $SU_v(2)$, the elements of which being
characterized by the 4-vector $\theta^\alpha$ with $\theta \cdot v=0$, as follows:
\begin{eqnarray}
&&H \to  e^{-i\,S_\alpha\,\theta^\alpha}\,H
\,, \qquad \quad
\bar H \to  \gamma_0\,(e^{-i\,S_\alpha\,\theta^\alpha}\,H)^\dagger\,\gamma_0 =
\bar H\,e^{+i\,S_\alpha\,\theta^\alpha}\,,
\nonumber\\
&& S^\alpha =\frac{1}{2}\,\gamma_5\, [\vslash, \gamma^\alpha] \,,
\qquad \;\;S^\dagger_\alpha \, \gamma_0=\gamma_0\,S_\alpha \,, \qquad \;\;
[\vslash, S_\alpha ]_- = 0 \,.
\label{spin-rotation}
\end{eqnarray}
Under a Lorentz transformation, characterized by the antisymmetric tensor $\omega_{\mu \nu}$,
the spinor part of the field transforms as
\begin{eqnarray}
&& H \to e^{i\,S_{\mu \nu}\,\omega^{\mu \nu}}\,H\,e^{-i\,S_{\mu \nu}\,\omega^{\mu \nu}}\,, \qquad \quad
\bar H \to e^{i\,S_{\mu \nu}\,\omega^{\mu \nu}}\,\bar H\,e^{-i\,S_{\mu \nu}\,\omega^{\mu \nu}}\,,
\nonumber\\
&& S_{\mu \nu} = \frac{i}{4}\,[\gamma_\mu, \gamma_\nu]\,.
\end{eqnarray}
It follows that only combinations of the form where the Dirac matrices are right of the field $H$ or left to the field
$\bar H$ are invariant under the spin group $SU_v(2)$. Now it is straightforward to construct the $SU_v(2)$-invariant effective Lagrange density bearing
the structures detailed in Eqs.~(\ref{def-La}-\ref{def-Lc}). We introduce
\begin{eqnarray}\label{LHQ}
{ \mathcal L} &=&-\,\frac{1}{2}\,{\rm tr } H   \Big\{ f^{(S)}_1(\bar B B)_+ +  f^{(S)}_2(\bar B B)_- + \frac 12 \, f^{(S)}_3{\rm tr} (\bar B B)
- f^{(S)}_4(\bar B_\alpha B^\alpha) - \frac 12  f^{(S)}_5 {\rm tr} (\bar B_\alpha  B^\alpha )  \Big\}\bar H \nonumber \\
&-& \,\frac{1}{2}\,{\rm tr } \, H \Big\{ f^{(A)}_1\,(\bar B\,\gamma_5\, \gamma^\mu\,B)_+ +  f^{(A)}_2\,(\bar B\,\gamma_5\, \gamma^\mu\,B)_-
 + \frac 12 \, f^{(A)}_3\,{\rm tr} \, (\bar B\,\gamma_5\, \gamma^\mu\,B) \Big\}  \,\gamma_5\,\gamma_\mu \,\bar H
\nonumber \\
&+& \,\frac{i}{4}\,{\rm tr } \,H  \Big\{ f^{(T)}_1\,(\bar B_\mu \, B_\nu) \,+
\frac 12 \, f^{(T)}_2\,{\rm tr}\, (\bar B_\mu \, B_\nu )
+ f^{(T)}_3\,(\bar B_\mu \, \gamma_5\,\gamma_\nu\,B -
\bar B \, \gamma_5\,\gamma_\nu\,B_\mu )
  \Big\}  \sigma^{\mu \nu }\,\bar H  \,.
\end{eqnarray}
We recall that the field $H$ is a three-dimensional row in flavor space, each of its components
consisting of a 4 dimensional Dirac matrix.
With the help of Eq.~(\ref{non-relativistic-expansion}) and Eq.~(\ref{phase-convention}) the chiral
and the $SU_v(2)$-invariant effective interactions can be expressed in terms of the fields $P_+$ and $P^\mu_+$.
By matching these expressions to each other we obtain 15 relations
\begin{align}\label{largeMc-SumRule}
c^{(S)}_1 &= \tilde c^{(S)}_1 = f_1^{(S)}\,, & c_2^{(S)} &= \tilde c_2^{(S)} = f^{(S)}_2\,,  & c_3^{(S)} &= \tilde c_3^{(S)} = f^{(S)}_3\,,
\nonumber\\
d_1^{(S)} &= \tilde d_1^{(S)} = f^{(S)}_4\,, & d_2^{(S)} &= \tilde d_2^{(S)} = f^{(S)}_5\,, \nonumber \\
c_1^{(A)} &= \tilde c^{(A)}_1 = f^{(A)}_1\,, & c^{(A)}_2 &= \tilde c^{(A)}_2 = f^{(A)}_2\,, & c^{(A)}_3 &= \tilde c^{(A)}_3 = f^{(A)}_3\,, \nonumber\\
d^{(E)}_1 &= \tilde d^{(E)}_1 = \tilde d^{(E)}_3= f^{(T)}_1\,, & d^{(E)}_2 &= \tilde d^{(E)}_2= \tilde d_4^{(E)} = f^{(T)}_2\,,\nonumber\\
e_1^{(A)} &= e_2^{(A)} = \tilde e^{(A)}_1= \tilde e^{(A)}_2 = f^{(T)}_3\,.
\end{align}
Having implemented the heavy-quark symmetry, out of 26 parameters from Section \ref{section:chiral-lagrangian} there remain 11 independent parameters only.

\section{Large-$N_c$ operator analysis }

In this section we further correlate the parameters of the effective interaction introduced in
Section \ref{section:chiral-lagrangian}. We follow the works of Luty and March-Russell \cite{Luty:Russel:1994} and  of Dashen, Jenkins and
Manohar \cite{Dashen:Jenkins:Manohar:1994}. These works introduced a formalism for a
systematic expansion of baryon matrix elements of QCD quark operators in powers of $1/N_c$. In a previous work two of the
authors detailed the application of this formalism to correlation functions involving a product of two axial-vector quark currents \cite{Lutz-Semke-2011}.
Here we further adapt the formalism for the case where charm changing quark currents occur. We consider
baryon matrix elements of the following products of two quark currents:
\begin{eqnarray}
&& C^{AA}_{\mu \nu,a} (q) = i\,\int d^4 x \,e^{-i\,q\cdot x}  \,{\mathcal T}\, A_\mu (0)\,\lambda_a\,\bar  A_\nu(x)\,,
\qquad \bar A_\mu(x) = A^\dagger_\mu(x)\,,
\nonumber\\
&& C^{VV}_{\mu \nu,a} (q) = i\,\int d^4 x \,e^{-i\,q\cdot x}  \,{\mathcal T}\; V_\mu (0)\,\lambda_a\,\bar V_\nu(x)\,,
\qquad \,\bar V_\mu(x) = V^\dagger_\mu(x)\,,
\nonumber\\
&& C^{VA}_{\mu \nu,a} (q) = i\,\int d^4 x \,e^{-i\,q\cdot x}  \,{\mathcal T}\, V_\mu (0)\,\lambda_a\, \bar A_\nu(x)\,,
\nonumber\\
&& \qquad \qquad \bar  A_\mu (x) = \bar c (x)\,\gamma_\mu \,\gamma_5\,
\left( \begin{array}{l}
u(x) \\
d(x) \\
s(x)
\end{array} \right) \,, \qquad  \bar V_\mu (x) = \bar c (x)\,\gamma_\mu \,\left( \begin{array}{l}
u(x) \\
d(x) \\
s(x)
\end{array} \right)\,,
\label{def-Cij}
\end{eqnarray}
with the quark field operators $u(x),d(x),s(x), c(x)$ of the up, down, strange and charm quarks. With $\lambda_a$ we
denote the Gell-Mann matrices supplemented with a singlet matrix $\lambda_0 = \sqrt{2/3}\,\one $.

In QCD the charm changing vector and axial-vector currents are accessed by means of classical source functions
${\mathcal V}^\mu(x)$ and ${\mathcal A}^\mu(x)$ that couple to the currents directly. Our effective Lagrangian
may be considered to be a function of such classical sources. All what we need in the following is the
existence of a direct coupling of these sources to the $D$ meson fields
\begin{eqnarray}
{\mathcal L} =  f_A \,\Big[ {\mathcal A}^\mu\,(\partial_\mu \,\bar D)  +(\partial_\mu \, D) \,\bar {\mathcal A}^\mu \Big]
+\frac{f_V}{M_V}\,\Big[ {\mathcal V}^\nu\,(\partial^\mu \,\bar D_{\mu\nu}) +(\partial^\mu \,D_{\mu\nu})\,\bar {\mathcal V}^\nu  \Big]\,,
\label{def-sources}
\end{eqnarray}
with some coupling constants $f_A$ and $f_V$. With Eq.~(\ref{def-sources}) the counter terms introduced in Section \ref{section:chiral-lagrangian}
contribute to the matrix elements of $C_{\mu \nu}^{(XY)}(q)$ ($X,Y = V,A$) in the baryon octet and decuplet states. According to
the LSZ reduction formalism, such contributions entail the corresponding contributions to the on-shell $D$ meson baryon
scattering amplitudes. They are identified unambiguously by taking the residuum of poles generated by the incoming
and outgoing $D$ meson. For the sake of more compact notation in what follows, it is convenient to introduce
\begin{eqnarray}
\bar C^{XY}_{\mu \nu,a} (\bar q, q) =
\frac{\bar q^2 - M^2_X}{f_X}\,C^{XY}_{\mu \nu, a} (q)\, \frac{q^2 - M^2_Y}{f_Y}\,,
\label{def-Cbar}
\end{eqnarray}
with $X,Y = V,A$ and $M_A $ and $M_V$ the masses of the pseudoscalar and vector $D$ mesons in the flavor $SU(3)$ limit.
In Eq.~(\ref{def-Cbar}) we identify $q_\mu$ and $\bar q_\mu$ with the 4-momenta of the incoming and outgoing $D$ mesons.

Our strategy is to first evaluate baryon matrix elements of the correlation function (\ref{def-Cbar}) with the
help of the chiral Lagrangian. Since each parameter in the chiral Lagrangian can be dialed freely
without violating any chiral Ward identity, we may focus on the contributions of the
two-body counter terms. Detailed results are collected in the Appendix.
In a second step we work out the $1/N_c$ expansion of the matrix elements of Eq.~(\ref{def-Cbar}).
Finally, we perform a matching of the two results. A correlation of the parameters of the chiral Lagrangian arises.

Following Ref.~\cite{Lutz-Semke-2011}, the baryon octet and decuplet states
\begin{eqnarray}
\ket{p, \,\chi,\, a }\,, \qquad \qquad \ket{p, \,\chi, \,ijk } \,,
\label{def-states}
\end{eqnarray}
are specified by the momentum $p$ and the flavor indices $a=1,\cdots ,8$ and $i,j,k=1,2,3$. The spin-polarization
label is $\chi = 1,2$ for the octet and $\chi =1,\cdots ,4$ for the decuplet states. At leading order in the $1/N_c$ the
baryon-octet and decuplet states  are degenerate. Thus, it is sometimes convenient to suppress the reference to the
particular state and use in the following $| p, \chi  \rangle $ as  a synonym for either a baryon-octet or baryon-decuplet
state.

The large-$N_c$ operator analysis rewrites matrix elements in the physical baryon states $\ket{p, \,\chi}$ in
terms of auxiliary states $\roundket{\chi}$  that reflect the spin and flavor structure of the baryons only. All
dynamical information is moved into effective operators. The $1/N_c$ expansion takes the generic form
\begin{eqnarray}
\bra{\bar p,\,\bar \chi} \, \bar C^{}_{\mu \nu,a}(\bar q,q)\, \ket{p,\,\chi} = \sum_n
c_n (\bar p, p) \roundbra{\bar \chi} \, O^{(n)}_{\mu \nu,a} \,\roundket{\chi}\,,
\label{def-expansion}
\end{eqnarray}
where we assume all 4-momenta to be on-shell. In the the center-of-mass frame the two three vectors
$\bar p$ and $p$ characterize the kinematics. The effective operators  $O^{(n)}_{\mu \nu,a}$ are
composed out of spin, flavor and spin-flavor operators, $J_i, T^a$ and $G^a_i$, which act on the
auxiliary states $\roundket{\chi}$. For $N_c=3$ we recall the results
of Refs.~\cite{Lutz:Kolomeitsev:2002,Lutz-Semke-2011}. It holds
\begin{eqnarray}
J_i \,\roundket{a,\chi}&=&\frac{1}{2}\, \sigma^{(i)}_{{\bar \chi} \chi}\, \roundket{a,{\bar \chi}}\,,
\nonumber \\
T^a\, \roundket{b,\chi}&=& i\,f_{bca}\, \roundket{c,\chi}\,,
\nonumber \\
G^{a}_i\, \roundket{b,\chi}&=&  \sigma^{(i)}_{{\bar \chi} \chi}\, \Big(\frac12\,d_{bca} + \frac{i}{3}\, f_{bca}\Big)\,
\roundket{c,{\bar \chi}} + \frac{1}{2\sqrt{2}}\, S^{(i)}_{{\bar \chi} \chi}\, \Lambda_{ab}^{klm}
\, \roundket{klm,{\bar \chi}}\,,
\nonumber \\
\nonumber \\
J_i \,\roundket{klm, \chi}&=& \frac{3}{2}\,\Big(\vec S \,\sigma_i\, \vec S^\dagger \Big)_{{\bar \chi} \chi}\,
\roundket{klm, {\bar \chi}},
\nonumber \\
T^a \,\roundket{klm, \chi}&=& \frac{3}{2}\,\Lambda^{a,nop}_{klm}\, \roundket{nop,\chi},
\nonumber \\
G^a_i \,\roundket{klm, \chi}&=& \frac34 \, \Big(\vec S\,\sigma_i\, \vec S^\dagger \Big)_{{\bar \chi} \chi}\,
\Lambda^{a,nop}_{klm}\, \roundket{nop, {\bar \chi}}  +
\frac{1}{2\sqrt{2}} \, \Big(S^{\dagger}_i \,\Big)_{{\bar \chi} \chi}\, \Lambda^{ab}_{klm}\, \roundket{b,{\bar \chi}}\,,
\label{result:one-body-operators}
\end{eqnarray}
with the Pauli matrices $\sigma_i$ and the transition matrices $S_i$ characterized by
\begin{eqnarray}
&& S^\dagger_i\, S_j= \delta_{ij} - \frac{1}{3}\sigma_i \sigma_j \,, \qquad
S_i\,\sigma_j - S_j\,\sigma_i = -i\,\varepsilon_{ijk} \,S_k\,,
\qquad \vec S\cdot   \vec S^\dagger= \one_{(4\times 4)}\,,
\nonumber\\
&& \vec S^\dagger \cdot  \vec S =2\, \one_{(2\times 2)}\,, \qquad \vec S \cdot \vec \sigma = 0 \,,\qquad
\epsilon_{ijk}\,S_i\,S^\dagger_j = i\,\vec S \,\sigma_k\,\vec S^\dagger\,.
\label{def:spin-transition-matrices}
\end{eqnarray}
In Eq.~(\ref{result:one-body-operators}) we introduced further the notation
\begin{eqnarray}
&& \Lambda_{ab}^{klm} = \Big[\varepsilon_{ijk}\, \lambda^{(a)}_{li}\,
\lambda^{(b)}_{mj} \,\Big]_{\mathrm{sym}(klm)}\,,\qquad \quad
\delta^{\,klm}_{\,nop} \;\;= \Big[\delta_{kn}\,\delta_{lo}\,\delta_{mp} \,\Big]_{\mathrm{sym}(nop)}\,,
\nonumber\\
&&  \Lambda^{ab}_{klm} = \Big[\varepsilon_{ijk}\, \lambda^{(a)}_{il}\,
\lambda^{(b)}_{jm} \,\Big]_{\mathrm{sym}(klm)}\,,\qquad \quad
\Lambda^{a,klm}_{nop} = \Big[\lambda^{(a)}_{kn} \delta_{lo} \,\delta_{mp}\Big]_{\mathrm{sym}(nop)}\,,
\label{def:flavor-transition-matrices}
\end{eqnarray}
which proved convenient in various derivations \cite{Lutz-Semke-2011}.
The symbol '$\mathrm{sym}(nop)$' in Eq.~(\ref{def:flavor-transition-matrices}) asks for a symmetrization
of the three indices $nop$, i.e. take the six permutations and divide out a factor $6$.

We return to the expansion (\ref{def-expansion}). There are infinitely many terms one may write down.
At a given order in the $1/N_c$ expansion a finite number of terms is relevant only. The counting is intricate since
there is a subtle balance of suppression and enhancement effects. An r-body operator consisting of the $r$ products
of any of the spin and flavor operators receives the suppression factor $N_c^{-r}$. On the other hand baryon matrix
elements taken at $N_c \neq 3$ are enhanced by factors of $N_c$ for the flavor and spin-flavor operators.
The counting advocated in \cite{Dashen:Jenkins:Manohar:1994}  may be summarized by the effective scaling laws
\begin{eqnarray}
J_i \sim \frac{1}{N_c} \,, \qquad \quad T^a \sim N^0_c \,, \qquad \quad G^a_i \sim N^0_c \,.
\label{effective-counting}
\end{eqnarray}
The counting rules (\ref{effective-counting}) by themselves are insufficient to arrive at significant results.
At a given order in the 1 over $N_c$ expansion there still is an infinite number of terms contributing. Taking higher
products
of flavor and spin-flavor operators does not reduce the $N_c$ scaling power. The systematic $1/N_c$ expansion is
implied by a set of operator identities \cite{Dashen:Jenkins:Manohar:1994,Lutz-Semke-2011}
which leads to the two reduction rules:
\begin{itemize}
\item All operator products in which two flavor indices are contracted using $\delta_{ab}$,
$f_{abc}$ or $d_{abc}$ or two spin indices on $G$'s are contracted using $\delta_{ij}$ or $\varepsilon_{ijk}$
can be eliminated.
\item All operator products in which two flavor indices are contracted using symmetric or antisymmetric
combinations of two different $d$ and/or $f$ symbols can be eliminated. The only exception to this rule is
the antisymmetric
combination $f_{acg}\,d_{bch}-f_{bcg}\,d_{ach}$.
\end{itemize}
As a consequence the infinite tower of spin-flavor operators truncates at any given order in the $1/N_c$ expansion.

We turn to the $1/N_c$ expansion of the baryon matrix elements of the operators in Eq.~(\ref{def-Cbar}).
We focus on the space components of the correlation functions.
In application of the operator reduction rules, the baryon matrix elements of the product of two quark currents are
expanded in powers of the effective
one-body operators according to Eq.~(\ref{def-expansion}). The ansatz for the momentum dependence of the expansion
coefficients in (\ref{def-expansion}) is furnished by the leading order terms in the corresponding low-energy expansion
stated in Eqs.~(\ref{result:matrix_elements_low_energy_expansion_AA}-\ref{result:matrix_elements_low_energy_expansion_AV}). In the course of the construction of the various structures, parity and time-reversal transformation properties are taken into account.
To the subleading order in the $1/N_c$-expansion we find the relevance of the following 11 effective operators
\begin{align}
\bra{\bar p, \bar \chi}  \,\bar C^{AA}_{ij,a}\, \ket{p, \chi}  & = \bar p_i\, p_j\,
\roundbra{\bar \chi} \, g^{AA}_1\, T^a
 + \smallfrac 12 \,g^{AA}_2\, \big[J_k, \,G_k^a\big]_+  \roundket{\chi},
\nonumber\\
\langle \bar p, \bar \chi |  \,\bar C^{VV}_{ij,a}\,| p, \chi  \rangle  &=
 \delta_{ij}\, ( \bar \chi | \, g^{VV}_1\,T^a + \smallfrac{1}{2}\,g^{VV}_2\, \big[J_k, \,G_k^a\big]_+
 \roundket{\chi} \nonumber \\
&+ i\, \epsilon_{ijk} \roundbra{\bar \chi }\, g^{VV}_3\, G^a_k + \smallfrac{1}{2}\,g^{VV}_4\, \big[J_k, \,T^a\big]_+
\roundket{\chi} \nonumber \\
&+ \roundbra{\bar \chi} \smallfrac{1}{2}\,g^{VV}_5\, \big[J_i, \,G_j^a\big]_+ +
\smallfrac{1}{2}\,g^{VV}_6\, \big[J_j, \,G_i^a\big]_+ \roundket{\chi},
\nonumber\\
\langle \bar p, \bar \chi |  \,\bar C^{VA}_{ij,a}\,| p, \chi  \rangle  & =
p_j\, ( \bar \chi | \, g^{VA}_1\,  G^a_i + \smallfrac{1}{2}\,g^{VA}_2\, \big[J_i, \,T^a\big]_+ +
\smallfrac{1}{2}\,g^{VA}_3\, i\, \varepsilon_{ikl}\, \big[J_k, \,G^a_l\big]_+ \roundket{\chi} .
\label{result:matrix_elements_Nc_expansion}
\end{align}

In contrast to Eq.~(\ref{result:one-body-operators}), where the flavor index $a$ takes the values $1, ...,8$,
the flavor index $a$ in Eq.~(\ref{result:matrix_elements_Nc_expansion}) runs from 0 to 8 with the
identification\footnote{The factor $1/\sqrt{6}$ in Eq.~(\ref{def-T0-G0}) follows from the normalization
of $\lambda_0$ and from the definition of the effective operators $T^a$ and $G^a_i$.}
\begin{eqnarray}
T^0 = \sqrt{\frac{1}{6}}\,\quarknumberoperator \,, \qquad \qquad G^0_i = \sqrt{\frac{1}{6}}\,J_i \,.
\label{def-T0-G0}
\end{eqnarray}
The matrix elements of the operators in Eq.~(\ref{result:matrix_elements_Nc_expansion}) are obtained by 
consecutive applications of the results in Eq.~(\ref{result:one-body-operators}). The matrix elements of all
symmetric combinations of two one-body effective operators can be found in the Appendix of Ref.~\cite{Lutz-Semke-2011}.

Combining the large $N_c$ and the low-energy expansions for the baryonic matrix elements of quark
operators (or products of them) under consideration, leads to the correlations between the unknown
parameters in both expansions. In our special case, matching of the three lines
in Eq.~(\ref{result:matrix_elements_Nc_expansion}) to the corresponding expressions in the Appendix
provides the following three sets of matching results:
\begin{eqnarray}
&& \!\!\!\!\! \!\!\!\!\! \!\!\!\!\! \!\!\! c^{(S)}_1 = \frac 38\, g_2^{AA}, \qquad
c^{(S)}_2 = \frac 12\, g_1^{AA} + \frac 14\,g_2^{AA} , \qquad
c^{(S)}_3 = g_1^{AA} - \frac 14\, g_2^{AA},
\nonumber \\
&& \!\!\!\!\! \!\!\!\!\! \!\!\!\!\! \!\!\! d^{(S)}_1 =  \frac 32 \,g_1^{AA} + \frac{15}{8}\, g_2^{AA} , \qquad
d^{(S)}_2 = 0\,,
\label{result:chiral_Nc_couplings_matching_AA}
\end{eqnarray}
and
\begin{eqnarray}
&& \tilde c_1^{(S)} = \frac 38\, g_2^{VV} + \frac 14 \, g_+^{VV} , \qquad
\tilde c_2^{(S)} =  \frac 12\, g_1^{VV} + \frac 14 \,g_2^{VV} + \frac 16\, g_+^{VV} , \qquad
\nonumber \\
&&  \tilde c_3^{(S)} = g_1^{VV} - \frac 14\, g_2^{VV} - \frac 16\, g_+^{VV},
\nonumber \\
&& \tilde c_1^{(A)} = - \frac 14\, g_3^{VV}, \qquad
\tilde c_2^{(A)} = - \frac 16\, g_3^{VV} - \frac 14\, g_4^{VV}, \qquad
\tilde c_3^{(A)} = \frac 16\, g_3^{VV} - \frac 12\, g_4^{VV} ,
\nonumber \\
&& \tilde d_1^{(S)} =  \frac 32\, g_1^{(VV}  +\frac{15}{8}\, g_2^{VV} + \frac 94\, g_+^{VV} , \qquad
\tilde d_2^{(S)} =0,
\nonumber \\
&& \tilde d_1^{(E)} =  \frac 32\,  g_3^{VV} + \frac 94\, g_4^{VV} -3\, g_+^{VV} , \qquad
\tilde d_2^{(E)} = \frac 32\,  g_4^{VV},
\nonumber \\
&& \tilde d_3^{(E)} =  \frac 32\, g_3^{VV} + \frac 94\, g_4^{VV} + 3\,g_+^{VV} , \qquad
\tilde d_4^{(E)} = \frac 32\,  g_4^{VV},
\nonumber \\
&&\tilde e_1^{(A)} = g_3^{VV} + \frac 32\,   g_-^{VV} - \frac 12\,  g_+^{VV} , \qquad
\tilde e_2^{(A)} =  g_3^{VV} +  \frac 32\, g_-^{VV} +  \frac 12\, g_+^{VV} ,
\label{result:chiral_Nc_couplings_matching_VV}
\end{eqnarray}
with $g_\pm^{VV} = \frac 12 \,(g_5^{VV} \pm g_6^{VV} )$, and
\begin{eqnarray}
&&c_1^{(A)} = \frac 14\, g_1^{VA}, \qquad
c_2^{(A)} = \frac 16\, g_1^{VA} + \frac 14 \, g_2^{VA},  \qquad
c_3^{(A)} = - \frac 16\, g_1^{VA} + \frac 12 \, g_2^{VA},
\nonumber \\
&&e_1^{(A)} = - g_1^{VA} + \frac 32\, g_3^{VA},
\nonumber \\
&&d_1^{(E)} = - \frac 32\, g_1^{VA} - \frac 94\, g_2^{VA} ,  \qquad d_2^{(E)} = - \frac 32\, g_2^{VA}.
\label{result:chiral_Nc_couplings_matching_VA}
\end{eqnarray}
Note that the parameter  $e_2^{(A)}$ does not occur in the presented results here.
The large-$N_c$ operator expansion for this parameter is based on the analysis of the matrix
elements of $\bar C^{AV}$, which may be decomposed in terms of additional parameters $g^{AV}_{1,2,3}$
in analogy to the decomposition of the matrix elements of $\bar C^{VA}$ in Eq.~(\ref{result:matrix_elements_Nc_expansion}).
The corresponding results follow from Eq.~(\ref{result:chiral_Nc_couplings_matching_VA})
by the replacements $e_1^{(A)}\to e_2^{(A)}$ and $g^{VA}_{1,2,3}\to g^{AV}_{1,2,3}$.  As a consequence it follows that
$g^{VA}_{1,2} =g^{AV}_{1,2}$, but not necessarily $g^{VA}_{3} =g^{AV}_{3}$.

Altogether we find 12 large-$N_c$ parameters relevant at leading order. Compared with the 26 chiral parameters
we expect a set of 14 sum rules. We group the sum rules into three parts
\begin{align}
c^{(S)}_3 &= 2\, \big(c^{(S)}_2 - c^{(S)}_1 \big),  &  d^{(S)}_1 &= 3\, \big( c^{(S)}_1 + c^{(S)}_2 \big),  &
d^{(S)}_2 &= 0, \nonumber \\
\hline
c_3^{(A)} &= 2\, \big( c_2^{(A)} - c_1^{(A)} \big), & d_1^{(E)} &= - 9\, c_2^{(A)},  &
d_2^{(E)} &= 2\, \big( 2\, c_1^{(A)} - 3\, c_2^{(A)} \big), \nonumber \\
\hline
\tilde c^{(S)}_3 &= 2\, \big(\tilde c^{(S)}_2 - \tilde c^{(S)}_1 \big),  & \tilde
d_1^{(S)} &= 3\, \big( \tilde c_1^{(S)} + \tilde c_2^{(S)} \big)  + \big( \tilde e_2^{(A)} - \tilde e_1^{(A)} \big),  &
\tilde d_2^{(S)} &= 0, \nonumber \\
\tilde c^{(A)}_3 &= 2\, \big(\tilde c^{(A)}_2 - \tilde c^{(A)}_1 \big),  &
\tilde d_1^{(E)} &= -9\, \tilde c_2^{(A)} - 3\, \big(\tilde e_2^{(A)} - \tilde e_1^{(A)} \big),  &
\tilde d_2^{(E)} &=\tilde d_4^{(E)} = 2\, \big( 2\, \tilde c_1^{(A)}  - 3\, \tilde c_2^{(A)} \big),  \nonumber \\
  & & \tilde d_3^{(E)} &= - 9\, \tilde c_2^{(A)} + 3\, \big(\tilde e_2^{(A)} - \tilde e_1^{(A)} \big),
\label{result:chiral_couplings_sum_rules}
\end{align}
where the first and the third parts correlate the coupling constants describing the interactions of
pseudoscalar and  vector $D$ mesons, respectively. The second part provides the analogous relations for the transition
interactions, i.e. terms with one pseudoscalar and one vector $D$ meson field.

We observe that given the third set of the sum rules in Eq.~(\ref{result:chiral_couplings_sum_rules}), the first
two parts are recovered by applying the results of the heavy-quark mass expansion as summarized
in Eq.~(\ref{largeMc-SumRule}). This is a remarkable result demonstrating the consistency of a combined heavy-quark and
large-$N_c$ expansion.

\section{Summary}

We derived sum rules for the leading order two-body counter terms of the chiral Lagrangian as
implied by a combined heavy-quark and large-$N_c$ analysis. There are altogether 26
independent terms in the chiral Lagrangian with baryon octet and decuplet fields that contribute to the $D$ and $D^*$
meson baryon scattering process at chiral order $Q^0$.

At leading order in the heavy-quark expansion we find the
relevance of 11 operators only. Additional sum rules were derived
from the $1/N_c$ expansion. Combining both expansions, the number of unknown parameters is further reduced to 5.
At present such sum rules can not be confronted directly with empirical information. They are useful constraints in
establishing a systematic coupled-channel effective field theory for $D$ meson baryon scattering  beyond the threshold
region.
\vskip0.3cm
{\bfseries{Acknowledgments}}
\vskip0.3cm
Daris Samart was supported by CHE-PhD-SW-SUPV from the
Commission on Higher Education, Thailand.

\newpage
\section{Appendix}

With the help of the effective interaction discussed in Section \ref{section:chiral-lagrangian} we evaluate the leading order contributions to the matrix elements of the correlator function in Eq.~(\ref{def-Cbar}) for the baryon-octet and baryon-decuplet states. As in the main part of the text, we focus on the space components only and specify the kinematics in the center of mass system.

The leading terms in the low-momentum expansion of the matrix elements of the product of two axial-vector currents are
\begin{align}
\bra{\bar p, \bar \chi, c}\, \bar C^{AA}_{ij,a}\,  \ket{p, \chi, b} & =  \bar p_i \, p_j\, \delta_{\bar \chi \chi}  \times \left\{
\begin{array}{r@{\;\;:\;\;}l}
  \left( 2 \,\sqrt{\frac 23} \,c^{(S)}_1 + \sqrt{\frac 32}\,c^{(S)}_3 \right) \delta^{bc} & a=0 \\
  2\, c^{(S)}_1 \,d^{abc} + 2\, c^{(S)}_2 \,i\,f^{abc} & a=1,\ldots 8,
\end{array}
\right. \nonumber \\
\bra{\bar p, \bar \chi, nop}\, \bar C^{AA}_{ij,a}\, \ket{p, \chi, b} &= 0, \nonumber \\
\bra{\bar p, \bar \chi, nop}\, \bar C^{AA}_{ij,a} \,\ket{p, \chi, klm} &=  \bar p_i \,p_j\, \delta_{\bar \chi \chi}    \times \left\{
\begin{array}{r}
  \left(\sqrt{\frac 23} \,d^{(S)}_1 + \sqrt{\frac 32} \, d^{(S)}_2 \right) \delta^{nop}_{klm} \\
  d^{(S)}_1 \,\Lambda^{a, rst}_{klm}\, \delta^{nop}_{rst}.
\end{array}
\right.
\label{result:matrix_elements_low_energy_expansion_AA}
\end{align}
Here and in the following the upper row corresponds to the singlet component of the correlation function, $\bar C_{ij, 0}$, whereas the second row specifies the matrix elements of its octet components with $a=1,\ldots 8$. Furthemore, the flavour summation indices are $r,s,t=1,2,3$.\\
The expansion for the product of two vector currents reads
\allowdisplaybreaks[1]
\begin{align}
\bra{\bar p, \bar \chi, c}\, \bar C^{VV}_{ij,a}\, \ket{p, \chi, b} & =
\delta_{ij}\, \delta_{\bar \chi \chi}  \times \left\{
\begin{array}{r}
  \left( 2 \,\sqrt{\frac 23} \,\tilde c^{(S)}_1 + \sqrt{\frac 32} \,\tilde c^{(S)}_3 \right) \delta^{bc}\\
  2\, \tilde c^{(S)}_1 d^{abc} + 2\, \tilde c^{(S)}_2 if^{abc}
\end{array}
\right. \nonumber \\
& - i\varepsilon_{ijk}\, \sigma^{(k)}_{\bar \chi \chi}  \times \left\{
\begin{array}{r}
  \left( 2 \,\sqrt{\frac 23}\, \tilde c^{(A)}_1 + \sqrt{\frac 32}\, \tilde c^{(A)}_3 \right) \delta^{bc}\\
  2\, \tilde c^{(A)}_1 d^{abc} + 2\, \tilde c^{(A)}_2 \,i\,f^{abc},
\end{array}
\right. \nonumber \\
\bra{\bar p, \bar \chi, nop}\, \bar C^{VV}_{ij,a}\, \ket{p, \chi, b} &=  - \frac{1}{2\,\sqrt 2}\,\big(S_i \sigma_j \big)_{\bar \chi \chi} \times
 \left\{
\begin{array}{r}
  0 \\
  \tilde e^{(A)}_1 \Lambda_{ab}^{nop}
\end{array}
\right. \nonumber \\
& + \frac{1}{2\,\sqrt 2}\,\big(S_j\, \sigma_i \big)_{\bar \chi \chi} \times
 \left\{
\begin{array}{r}
  0 \\
  \tilde e^{(A)}_2 \,\Lambda_{ab}^{nop},
\end{array}
\right. \nonumber \\
\bra{\bar p, \bar \chi, nop}\, \bar C^{VV}_{ij,a}\, \ket{p, \chi, klm} &=
\delta_{ij}\, \delta_{\bar \chi \chi} \times \left\{
\begin{array}{r}
  \left(\sqrt{\frac 23}\,\tilde  d^{(S)}_1 + \sqrt{\frac 32} \, \tilde d^{(S)}_2 \right) \delta^{nop}_{klm}\\
  \tilde d^{(S)}_1 \,\Lambda^{a, rst}_{klm} \,\delta^{nop}_{rst}
\end{array}
\right. \nonumber \\
&  + \frac 12 \,\big(S_i S^\dagger_j \big)_{\bar \chi \chi}  \times \left\{
\begin{array}{r}
  \left(\sqrt{\frac 23} \,\tilde d^{(E)}_1 + \sqrt{\frac 32} \, \tilde d^{(E)}_2 \right) \delta^{nop}_{klm}\\
  \tilde d^{(E)}_1 \Lambda^{a, rst}_{klm} \,\delta^{nop}_{rst},
\end{array}
\right. \nonumber \\
& - \frac 12\, \big(S_j \,S^\dagger_i \big)_{\bar \chi \chi} \times \left\{
\begin{array}{r}
  \left(\sqrt{\frac 23}\, \tilde d^{(E)}_3 + \sqrt{\frac 32}\,  \tilde d^{(E)}_4 \right) \delta^{nop}_{klm}\\
  \tilde d^{(E)}_3\, \Lambda^{a, rst}_{klm} \,\delta^{nop}_{rst}.
\end{array}
\right.
\label{result:matrix_elements_low_energy_expansion_VV}
\end{align}
Finally, for the product of a vector and an axialvector currents we obtain:
\begin{align}
\bra{\bar p, \bar \chi, c}\, \bar C^{VA}_{ij,a}\, \ket{p, \chi, b} & =
  p_j\, \sigma^{(i)}_{\bar \chi \chi}  \times \left\{
\begin{array}{r}
  \left( 2 \,\sqrt{\frac 23} c^{(A)}_1 + \sqrt{\frac 32} \,c^{(A)}_3 \right) \delta^{bc}\\
  2\, c^{(A)}_1 \,d^{abc} + 2\, c^{(A)}_2 \,i\,f^{abc},
\end{array}
\right. \nonumber \\
\bra{\bar p, \bar \chi, nop}\, \bar C^{VA}_{ij,a}\, \ket{p, \chi, b} &= - \frac{1}{2\,\sqrt 2}\, p_j\, S^{(i)}_{\bar \chi \chi} \times
 \left\{
\begin{array}{r}
  0 \\
  e^{(A)}_1 \Lambda_{ab}^{nop},
\end{array}
\right. \nonumber \\
\bra{\bar p, \bar \chi, nop}\, \bar C^{VA}_{ij,a}\, \ket{p, \chi, klm} &=
-\frac 12\, p_j\, \big(\vec S \sigma^{(i)}\, \vec S^\dagger \big)_{\bar \chi \chi} \times \left\{
\begin{array}{r}
  \left(\sqrt{\frac 23}\, d^{(E)}_1 + \sqrt{\frac 32} \, d^{(E)}_2 \right) \delta^{nop}_{klm}\\
  d^{(E)}_1 \Lambda^{a, rst}_{klm}\, \delta^{nop}_{rst}.
\end{array}
\right.
\label{result:matrix_elements_low_energy_expansion_AV}
\end{align}
The spin-transition matrices $\vec \sigma$ and $\vec S$  and the flavor transition tensors $\delta_{klm}^{nop}$, $\Lambda^{nop}_{ab}$ and $\Lambda^{c,nop}_{klm}$  used above, were already introduced in Eq.~(\ref{def:spin-transition-matrices}) and Eq.~(\ref{def:flavor-transition-matrices}), respectively.

\end{document}